\documentclass[a4paper,fleqn,usenatbib]{mnras}

\usepackage[T1]{fontenc}
\usepackage{ae,aecompl}

\usepackage{graphicx}	
\usepackage{amsmath}	
\usepackage{amssymb}	

\usepackage{wasysym} 

\newcommand\newt{{\rm (newt)}}

\usepackage[usenames,dvipsnames,svgnames,table]{xcolor}

\usepackage[normalem]{ulem} 

\title[Multi-species initial conditions]{Accurate initial conditions in mixed Dark Matter--Baryon simulations}

\author[Valkenburg \& Villaescusa-Navarro]{
Wessel Valkenburg,$^{1}$\thanks{E-mail: wessel.valkenburg@cern.ch}
Francisco Villaescusa-Navarro$^{2,3,4}$\thanks{E-mail: fvillaescusa@simonsfoundation.org}
\\
$^{1}$Theoretical Physics Department, CERN, CH-1211 Gen\'eve 23, Switzerland\\
$^2$Center for Computational Astrophysics, 160 5th Ave, New York, NY, 10010, USA\\
$^3$INAF, Osservatorio Astronomico di Trieste, via Tiepolo 11, I-34131 Trieste, Italy\\
$^4$INFN - National Institute for Nuclear Physics, Via Valerio 2, I-34127 Trieste, Italy
}

\begin{document}
\label{firstpage}
\pagerange{\pageref{firstpage}--\pageref{lastpage}}
\maketitle

\begin{abstract}
We quantify the error in the results of mixed baryon--dark-matter hydrodynamic simulations, stemming from outdated approximations for the generation of initial conditions. The error at redshift 0 in contemporary large simulations, is of the order of few to ten percent in the power spectra of baryons and dark matter, and their combined total-matter power spectrum. After describing how to properly assign initial displacements and peculiar velocities to multiple species, we review several approximations: (1) {using the total-matter power spectrum to compute displacements and peculiar velocities of both fluids}, (2) scaling the linear redshift-zero power spectrum back to the initial power spectrum using the Newtonian growth factor ignoring homogeneous radiation, (3) using longitudinal-gauge velocities with synchronous-gauge densities, and (4) ignoring the phase-difference in the Fourier modes for the offset baryon grid, relative to the dark-matter grid. 
{
Three of these approximations do not take into account that dark matter and baryons experience a scale-dependent growth after photon decoupling, which results in directions of velocity which are {\em not} the same as their direction of displacement.
}
We compare the outcome of {hydrodynamic} simulations with these four approximations to our reference simulation, all setup with the same random seed and simulated using {\sc Gadget-III}. 
\end{abstract}


\section{Introduction}
Upcoming large-scale surveys such as LSST\footnote{https://www.lsst.org/}, Euclid\footnote{http://www.euclid-ec.org/} and SKA\footnote{https://www.skatelescope.org/} will map the three-dimensional spatial distribution of galaxies and cosmic neutral hydrogen with an unprecedented precision over very large cosmological volumes. The goal of these surveys is to tightly constrain the values of the parameters in the current cosmological paradigm, and thus to learn about the nature of dark energy, the geometry of the Universe and the nature of the initial conditions of the Universe.

The constraints on the cosmological parameters are derived by comparing cosmological observables, such as the shape of the power spectrum or the position of the Baryon Acoustic Oscillation (BAO) peak against predictions from theory. Hence, in order to decrease the error on the parameters, two ingredients are needed: precise cosmological observable measurements and accurate theoretical predictions.

Theoretical predictions at the linear order are extremely accurate. For instance, codes such as {\sc CAMB} \citep{CAMB} or {\sc CLASS} \citep{CLASS} can compute the linear matter power spectrum to a precision of $0.01\%$ \citep{CLASS_vs_CAMB}, by numerically integrating the hierarchy of the Boltzmann equations. Unfortunately, {while} linear theory is valid at all scales only in the early Universe, {in the late Universe its validity is limited to the} largest scales. For instance, {non-linear gravitational evolution induces a damping and broadening on the BAO peak \citep{Crocce_2008,Padmanabhan_2009,Baldauf_2015,peloso,Blas:2016sfa} that takes place on very large scales ($\sim100~h^{-1}$Mpc).} 

Having accurate theoretical predictions of cosmological observables in the mildly or fully non-linear regime is one of the main topics of current cosmological research. The goal is to maximally profit from the aforementioned future observations in these regimes, and leave no information unexploited. 

For example, a falsification of the cosmological constant as the sole force driving the accelerated expansion of the Universe, would have far-going consequences for our understanding of physics. Moreover, a deeper understanding of the formation of structure in the Universe, can shed light on gravity, the nature of dark matter {and the mass of neutrinos}. Thus, the constraints on the cosmological parameters can be largely tightened by employing the information contained in that regime. 

There are different ways to approach the mildly non-linear regime. The most important ones are perturbation theory and numerical simulations. Plain perturbation theory \citep{PT_review} has proven as a very powerful tool to study the mildly non-linear regime as is currently employed in cosmological analysis \citep[see e.g.][]{Sanchez_2016, Salazar_16,Alam_16,Chuang_16}. Unfortunately, perturbation theory will break down when the amplitude {of any perturbation in the system becomes large.} 

Attempts to go beyond plain perturbation theory are made, under the common name of Effective Field Theories (EFTs)~\citep{Baumann:2010tm,Anselmi:2012cn,Blas:2015tla,Floerchinger:2016hja, Lewandowski_15, Foreman_15,Carrasco_12, Pietroni_14}. An EFT allows one to parametrise the effect of the unknown (or at least uncomputable) physics of, in this case, the small highly nonlinear scales, on the scales at study in the mildly nonlinear regime {and vice versa}. In EFTs, the perturbations on large scales are then no longer described by perturbations in a pure dark-matter fluid, but by perturbations on a fluid endowed with a sound speed, equation of state, bulk and shear viscosities, and higher order parameters. The values of these parameters need to be gauged using the results of fully nonlinear simulations. But once known, they can be used to perform higher order perturbation theory on the effective fluid, thus slightly enlarging the regime of validity of perturbation theory. EFT is then hoped to be useful for decreasing the number of simulations necessary for the exploration of parameter space. Nevertheless, for EFT to be useful, simulations must first reach the desired accuracy.

Under all other circumstances, i.e. in the fully non-linear regime, theoretical predictions are limited to the output of numerical simulations. 
Numerical simulations are one of the most, if not the most, powerful tools available to study cosmology, since they provide information on all regimes (from linear to fully non-linear) and in all the ingredients involved in relating theory with observations: matter power spectrum, redshift space distortions and galaxy/halo bias. Cosmological numerical simulations can be classified in two broad categories: 1) N-body simulations, where the evolution of a set of collisionless particles is numerically integrated and 2) hydrodynamic simulations, where the gravitational evolution is solved together with the equations governing the hydrodynamic forces affecting baryons. {Usually, hydrodynamic simulations include an N-body simulation to model the Dark Matter dynamics in addition to the hydrodynamics.}

The output of numerical simulations depends on several factors such as the accuracy of the integrator, the method used to compute forces, the way the initial conditions are generated, the volume simulated and so on. The method and value of the parameters controlling these ingredients have to be wisely chosen given the accuracy required for the simulation output. A comparison among the output of different codes, the impact of mass resolution and volume effects has been recently investigated in \cite{Schneider_2015}. The purpose of the this paper is to study the impact of various simplifications used when generating the initial conditions, paying particular attention to the case of hydrodynamic simulations. In particular, we will identify systematic errors commonly committed when generating the initial conditions and its impact in terms of cosmological observables. 

{Our focus lies on the connection between linear theoretical transfer functions, and the initial density and velocity fields of baryons and dark matter {in numerical simulations}.  The issue of initial conditions for multiple-species was addressed by some references. \cite{Yoshida:2003sy} focused on much smaller scales, not relevant for Baryon Acoustic Oscillations. \cite{Angulo:2013qp} adapted velocities computed in the longitudinal gauge, an approximation which later was pointed out not to be entirely correct (see the body of the paper for a clarification and quantification of this statement). At last, \cite{Valkenburg:2015dsa} focused on scale dependence in scalar fields, paying no attention to baryons. } We do not focus on the issue of resolution, for which we direct the reader to for example~\cite{Gabrielli:2001xw,Garrison:2016vvp}.

This paper is organised as follows. In section \ref{sec:ICs} we discuss the Gauge in which initial conditions are generally generated and the impact of relativistic effects on them. We discuss the different approximations people usually employ when creating the initial conditions in section \ref{sec:approximations}.  {In sections \ref{sec:errors_ICs} and \ref{sec:errors} we quantify the errors induced by employing the different approximations.} Finally, we draw the main conclusions of this work in section \ref{sec:conclusions}.

\section{Consistent initial conditions}
\label{sec:ICs}

\subsection{Relativistic versus Newtonian gravity}
Apart from a few exceptions~\citep{Adamek:2016zes,Giblin:2015vwq,Bentivegna:2015flc}, cosmological simulations are predominantly done using Newtonian gravity~\citep{Kuhlen:2012ft}. Relativistic effects can occur at two ends of the simulation spectrum: near black holes (small scales) and when scales become comparable to and larger than the Hubble radius (large scales). In order for a Newtonian simulation to be an accurate realization of the cosmological model, black holes must form only below the resolution of the simulation, and Hubble radius effects must be under control. 

The former assumption may be questioned, as mass estimates of the largest observed (possible) black holes range up to $\mathcal{O}(10^{10}) M_{\astrosun}$~\citep{2016Natur.532..340T}, which contain an amount of mass which originates from a primordial co-moving volume of $\mathcal{O}(1)$ Mpc$^3$. This question is however not part of this work.

Let us summarize why the latter assumption, that relativistic effects at scales beyond the Hubble radius are under control, is correct.

Formally, when any scale becomes nonlinear, the entire linear perturbation theory is invalidated, as one has no control over interaction between short and large scale modes. However, as argued in~\cite{1974A&A....32..391P}, small, random, momentum conserving displacements introduce additional power $\propto k^4$ in the power spectrum~\footnote{To repeat the argument of~\cite{1974A&A....32..391P} in brief: a random small displacement of masses on short scales implies a differentiation of the density field, effectively introducing a factor of $k$ in Fourier space. A random displacement that conserves momentum however, such as one caused by gravitational dynamics, lets the displacements of the masses cancel in pairs, such that it only introduces a second derivative, $k^2$, in the density field, and hence introduces additional power $\propto k^4$ in the power spectrum}. As long as the power spectrum of perturbations scales as $k^n$ with $n<4$, which is the case in the current paradigm { \citep[$n_s\cong0.96$,][]{Planck_15}}, primordial power on large scales will dominate by orders of magnitude over newly introduced power coming from small-scale nonlinear dynamics. 

The argument is confirmed by simulations: the transition from linear to nonlinear dynamics is hierarchical, moving from small scales to large scales. This means that at any time one can always go to large enough scales to find that linear perturbation theory still gives the correct description there. This holds in both Newtonian and relativistic simulations~\citep{Adamek:2016zes}.

Vice versa, small scale nonlinear modes obviously are coupled to the large scale linear modes (hence the term nonlinear). Since the nonlinear effects are large, one may wonder what the effect of a change in large-scale perturbation theory would be on small scale modes. This worry may be fed by the observation that relativistic perturbation theory may show different behaviour on super-Hubble scales than on sub-Hubble scales, a distinction that is absent in Newtonian gravity. However, in the following sections we argue why this worry is unfounded when one considers pressureless species only, such as dark matter and baryons (after photon decoupling). In brief, Newtonian and relativistic gravity provide the same large-scale linear behaviour for pressureless matter when there are no perturbations in relativistic species such as photons and neutrinos. Hence, the coupling between small- and large-scale modes is taken into account correctly in Newtonian simulations. Super-hubble effects on perturbations in pressureless matter are a gauge artefact, which we clarify in sections~\ref{sec:nbgauge} and \ref{sec:horizons}.

\subsection{N-body gauge in brief}\label{sec:nbgauge}

The Friedman-Lema\^itre-Robertson-Walker metric in conformal time endowed with scalar perturbations only, in an arbitrary gauge can be written as:
\begin{align}
\frac{ds^2}{a(\tau)^2}=&-(1+2A) d\tau^2 - 2 B_i d\tau\,dx^i \nonumber\\
&+ \left[(1+2H_L)\eta_{ij} + 2h^T_{ij}  \right]dx^idx^j ,\label{eq:anyGauge}
\end{align}
with the Minkowsky metric $\eta_{\mu\nu}$, and
\begin{align}
B_i =& \int \frac{d^3k}{\left(2\pi\right)^3}\frac{k_i}{k}B_{{\vec k}} e^{i\vec k \cdot \vec x},\\
h^T_{ij} =& \left[\frac{\partial_i\partial_j}{\nabla^2} - \tfrac{1}{3}\eta_{ij}\right]H_T.
\end{align}
The dimensionless $A$, $B$, $H_L$ and $H_T$ are small compared to $1$, and depend on the coordinates.

The perturbations in a pressureless fluid in an arbitrary gauge are,
\begin{align}
	\dot{\bar\rho} =& - 3\mathcal{H}\bar\rho,\label{eq:bgDensity}\\
\dot\Delta_\rho =&  - \theta - 3\dot  H_L,\label{eq:energyConservation}\\
\dot \theta + k \dot B =&  k^2 A    - \mathcal{H} (\theta + kB ),\label{eq:velocityEquation}
\end{align}
where {$\Delta_\rho\equiv\delta\rho/\bar\rho$}, $\mathcal{H}\equiv \dot a(\tau)/a(\tau)$, $\dot{\,\,} \equiv d/d\tau$ {and $\tau$ is the conformal time}. Both dark matter and baryons satisfy these equations at the redshifts where initial conditions for simulations are typically set, and only differ in their initial values for density $\rho$ and {peculiar} velocity $\theta = \vec\nabla \cdot \vec u$. Note that at this point we are only considering linear perturbations after photon decoupling, in which case baryons can be described as a pressureless fluid, just like dark matter.

The Newtonian equivalent of these equations are,
\begin{align}
	\dot{\bar\rho}^{\newt} =& - 3\mathcal{H}\bar\rho^{\newt},\label{eq:bgDensityNewt}\\
\dot\Delta_\rho^{\newt} =&  - \theta^{\newt} ,\label{eq:energyConservationNewt}\\
\dot \theta^{\newt}  =&  k^2 A    - \mathcal{H} \theta^{\newt} ,\label{eq:velocityEquationNewt}
\end{align}
where $A$ is identified with the Newtonian potential. 

A gauge can be chosen where in the presence of solely a single pressureless matter species, Eqs.~(\ref{eq:bgDensity}--\ref{eq:velocityEquation}) and Eqs.~(\ref{eq:bgDensityNewt}--\ref{eq:velocityEquationNewt}) match, {\em and} where the potential $A$ satisfies the Poisson equation, $-k^2 A = 4\pi G_{\rm N} \bar \rho a^2 \delta$, the N-body gauge~\citep{Fidler:2015npa} (see also~\cite{Flender:2012nq}). {In the N-body gauge, the full set of equations (\ref{eq:bgDensity}--\ref{eq:velocityEquation}) become identical to equations (\ref{eq:bgDensityNewt}--\ref{eq:velocityEquationNewt}), both completed by the same Poisson equation. Clearly, the implication is that at the linear level and in the presence of only pressureless matter, Newtonian gravity solvers correctly give the general relativistic solution.}

Beware that this solution deviates from the old standard lore of either (a) identifying the longitudinal gauge with that of Newtonian gravity, or (b) combining peculiar velocities from the longitudinal gauge with densities from the synchronous co-moving gauge, as was done previously~\citep[e.g.][]{Chisari:2011iq,Rigopoulos:2014rqa,Hahn:2016roq}. 

As explained in~\cite{Fidler:2015npa}, the matter density spectra $\left<\rho_{k}^2\right>$ in the N-body gauge are equal to that of any co-moving gauge, e.g. the matter density spectra in the synchronous co-moving gauge. It is then straightforward to conclude that velocities can be obtained easily and consistently by taking the time derivative of synchronous gauge density perturbations, as readily obtainable from standard Boltzmann solvers {such as} {\sc Class} \citep{CLASS} and {\sc Camb} \citep{CAMB}.

\subsection{Horizons and Newtonian simulations}\label{sec:horizons}
It is common lore to speak of the Hubble radius as a horizon, and to think differently of super- and sub-Hubble modes in linear perturbation theory. Let us reiterate some arguments that can be found in~\cite{Sawicki:2012re}, in order to alleviate any misunderstanding. 
First of all, the cosmic horizon is not equal to the Hubble radius; super-Hubble modes grow just as happily as sub-Hubble modes in the late universe in the standard cosmological model. Gravity acts beyond the Hubble radius. Even stronger, in absence of relativistic species, such as in a pure dust Universe with an optional cosmological constant, the growth on super- and sub-Hubble scales is the same. This is no longer the case when neutrino and photon perturbations are taken into account.

{There are possibly two scales that can play a role in the perturbations in a single fluid: (1) the Compton scale, which roughly speaking determines the transition from a classical to quantum mechanical description, and which we will ignore here, and (2) the Jeans scale or sound horizon, which is set by the fluid's sound speed. On scales below the Jeans scale, the perturbations are affected by the fluid's pressure. On scales above the Jeans scale, the perturbations in the fluid behave by all means as if it were pressureless: dust. Note that the background expansion {\em does} depend on the fluid's pressure, on all scales.}

{For a pressureless fluid, the sound speed and hence Jeans length are zero, which is reflected by the scale independence of the growth of Dark Matter perturbations in any co-moving gauge. This proves that the Hubble radius plays no role in the perturbations. {The statement is reinforced by the existence of the N-body gauge, as explained in the previous section.} The scale dependent growth of dust perturbations in the Longitudinal gauge, is truly a gauge artefact.}

{During inflation on the other hand, in the most common models, the sound speed of perturbations is equal or close to the speed of light, such that the scale that enters in perturbation equations for scalar perturbations is linearly related to that in (geometrical) tensor perturbations{, which also propagate at the speed of light}. The Jeans length corresponding to a sound speed equal to the speed of light, {\em is} proportional to the Hubble parameter.}

{When photons and baryons are still tightly coupled, the sound speed in that coupled matter is one third of the speed of light, such that the sound horizon (Jeans length) {\em does} behave proportional to the Hubble radius, but its effect has nothing to do with General Relativity and the presence of a particle horizon.}

{A common method used to generate initial conditions for numerical simulations consists in computing the $z=0$ linear power spectrum through {\sc CAMB} or {\sc CLASS} and "rescale" its amplitude to the starting redshift of the simulation. The "rescaling" is carried out by using the same Newtonian physics that the simulation follows, which translates into a scale-independent growth for the $\Lambda$CDM model (see \cite{Zennaro:2016nqo} for the rescaling implementation in massive neutrino cosmologies). This method guaranties that the power spectrum from a Newtonian simulation at low redshift on linear scales will agree with the output of Boltzmann solvers, while it will differ at high redshifts.} An often heard argument for this approximation, is that a Newtonian simulation does not correctly model super-Hubble effects, because it has no knowledge of General Relativity. As per the N-body gauge and the above arguments, this is a misunderstanding: the Hubble radius and its effect on linear dust perturbations, are only present in particular coordinate gauges (such as the longitudinal gauge), which are simply {\em not} the coordinate gauge in which a Newtonian simulation acts. The growth factor used for this re-scaling, in retrospect, is relativistic as it is identical to the growth factor in the N-body gauge. {What is wrong about this approach though, is that it treats Baryons and Dark Matter as if they were the same single species.}

Regardless of horizons however, what is indeed a mismatch between Newtonian perturbations modeled in a simulation, and the linear perturbation theory solved for in a relativistic Boltzmann code such as {\sc Camb} and {\sc Class}, is the effect of perturbations in relativistic species. Newtonian simulations (apart from~\cite{Brandbyge:2016raj}) do not take the $1\%$ contribution to the amplitude of perturbations in the gravitational potential into account, as argued in subsection~\ref{subsec:radiationpert}. This is the only valid argument for applying the rescaling: {$T_i(\tau(z_i), \vec k) =  D_{\rm N}(z_i)/D_{\rm N}(z=0) T_i(\tau(z=0), \vec k)$, where $z_i$ is the simulation starting redshift.} {Still, the rescaling needs to properly take into account the physical model that is simulated; each of the species may have its own transfer function (growth factor), depending on the simulation is will be run~\citep{Yoshida:2003sy,Angulo:2013qp,Zennaro:2016nqo}.}

The aim of this paper is exactly to show that ignoring the relative difference between baryons and dark matter, is of much greater importance than the  $\sim1\%$ effect (limited to very large-scales) of ignoring the perturbations arising from relativistic species.

\subsection{Multiple species}
The N-body gauge reduces to Newtonian perturbation theory {\em only when a single pressureless species is present.}

\subsubsection{Baryons and Dark Matter}
Provided that all perturbations are adiabatic (that is, there are no iso-curvature perturbations), baryons and dark matter (and all other species, for that matter) have identical power spectra, and behave as a truly single pressureless species~\citep{Sawicki:2012re}. Therefore, no relativistic corrections on {super-hubble} scales are to be suspected. 

Deep inside the Hubble radius, Newtonian perturbation theory is anyway known to be in agreement with relativistic perturbation theory~\citep{PT_review}, because velocities are small compared to the speed of light. The evolution of both species differs from a simple single species (total matter) scenario, at linear and higher order~\citep{Bernardeau:2011vy,Somogyi:2009mh}, the linear part of which is correctly described by the synchronous-gauge density spectrum and its time derivative. {In other words, both Newtonian and relativistic perturbation theory agree on the relation between single- and multi-species evolution on sub-Hubble scales.}

Precisely at the Hubble radius, a difference between Newtonian and relativistic perturbation theory could arise {from} a large difference in the velocities of baryons and dark matter. This difference is sourced by the coupling between photons and baryons, which, if present, anyway invalidates Newtonian perturbation theory. Hence, if initial conditions for a simulation are set sufficiently far after photon decoupling (which happens at a redshift of $z \simeq 1100$), such as for example $z_i=127$ as in this paper, there will be no error: one can safely endow species with the relativistic power spectrum computed in the synchronous gauge, and with velocities that are determined by the time derivative of the synchronous-gauge density power spectrum.

\subsubsection{Ignoring perturbations in neutrinos and photons}\label{subsec:radiationpert}
 The power spectra that we use for the generation of initial conditions, are computed using the Boltzmann code {\sc Class}, for a cosmology which includes perturbations in photons and neutrinos. These are absent in the N-body gauge. Using {\sc Class} one can easily verify that these relativistic species contribute up to 1\% in power in the gravitational potential at redshift $z=127$. Their (linear) contribution however quickly decays. Since their contribution is much less than that of baryons~\citep[e.g.][]{Palanque_2015,Cuesta_2016}, {the error from incorrectly taking their contribution into account should be much less than that of providing baryons the wrong initial spectrum, however still of the order of percents~\citep{Brandbyge:2016raj}.} Moreover, the relativistic shear introduced by massive neutrinos has been found to be undetectable, when comparing relativistic~\citep{Adamek:2016zes} to Newtonian~\citep{Zennaro:2016nqo} simulations. 
Nevertheless, the fact that we make the same systematic approximation in the various approaches that we test, means that our qualitative conclusions about baryons relative to dark matter probably are robust.

\section{Various approximations in the literature}
\label{sec:approximations}

Initial conditions that carry adiabatic perturbations for multiple species, are generated by first setting up a single source density field $\rho_{\vec k, {\rm ini}}$ ~\citep{Angulo:2013qp,Valkenburg:2015dsa}, and convolving with the transfer function for each species $i$, 
\begin{align}
	\delta\vec x_i &= -L^{-\frac{3}{2}}\sum_{\vec k=-\infty}^{\infty}  \frac{i\vec k}{k^2} T_i(\tau, \vec k)\rho_{\vec k, {\rm ini}} e^{i\vec k\cdot \vec x},\\
		\vec v_{i} = \frac{d}{d\tau}\delta\vec x_i &= -L^{-\frac{3}{2}}\sum_{\vec k=-\infty}^{\infty}  \frac{i\vec k}{k^2} \frac{d}{d\tau}T_i(\tau, \vec k)\rho_{\vec k, {\rm ini}} e^{i\vec k\cdot \vec x}.
\end{align}

Our reference approach, consists of taking the synchronous gauge transfer function $T_i(\tau, \vec k)$ of each specific species from {\sc Class} at redshift $z=127$. The simplicity of the correct approach leaves no reason for any of the following approximations, the errors of which are under study here.

\subsection{All as total matter}\label{subsec:multispecies}
\begin{figure}
\begin{center}
{\large Linear growth rate at z = 127}
\includegraphics[width=\columnwidth]{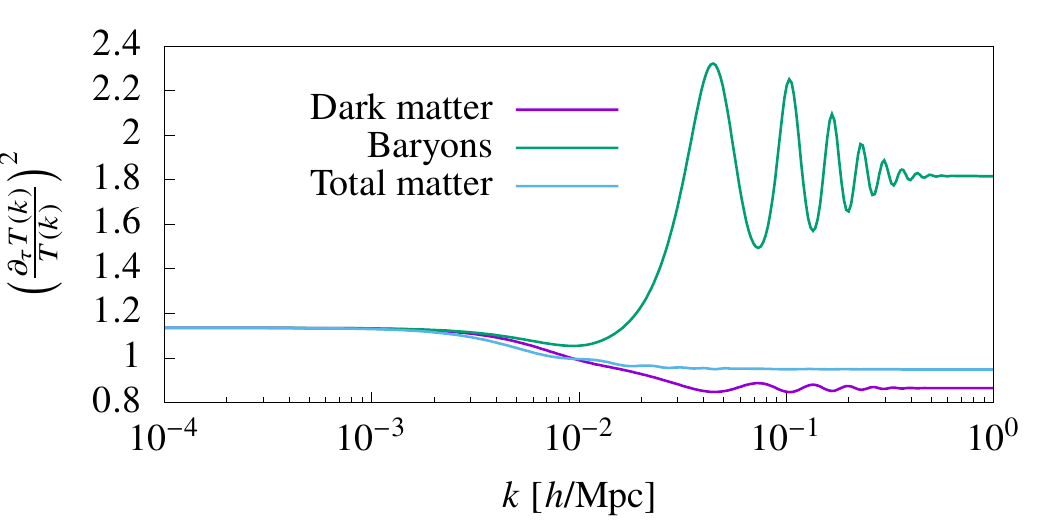}
\setlength{\unitlength}{0.0500bp}%
\end{center}
\vspace*{0mm}
\caption{Linear growth rate of baryons (green), dark matter (magenta) and total matter (blue) at $z=127$ from {\sc CLASS} for the cosmological model considered in this work. While the growth rate of total matter is scale-independent on scales below the horizon, the growth rates of dark matter and baryons in particular, exhibit a significant scale-dependence. \label{fig:growth_rate}}
\end{figure}

The first and oldest approach, is to give both dark matter and baryons the power spectrum of total matter, which is a weighted sum of {the} baryons {($T_{\rm b}$)} and dark matter {($T_{\rm dm}$)} {transfer functions}: 
\begin{equation}
T_{\rm m}=\left(\frac{\Omega_{\rm b}}{\Omega_{\rm m}}\right)T_{\rm b}+\left(\frac{\Omega_{\rm dm}}{\Omega_{\rm m}}\right)T_{\rm dm}~,
\end{equation}
where $\Omega_{\rm m}=\Omega_{\rm b}+\Omega_{\rm dm}$. In this case, the initial displacements are computed using the above power spectrum for each species and the peculiar velocities are assigned using the scale-independent growth of total matter (see Fig. \ref{fig:growth_rate}).

\subsection{Rescaling P(k, z=0), ignoring radiation at the background level}\label{subsec:veryoldway}
Often simulations take only Newtonian matter species (dark matter and baryons) and a cosmological constant into account when computing the Hubble factor, which enters the dynamics as an effective friction on the particles motion. The detailed evolution of the linear power spectrum is slightly different in a Universe without radiation, as it does not show the small scale-dependent growth near the Hubble radius that is present in a Universe with a homogeneous radiation component. It is therefore custom to take the redshift $z=0$ output of a Boltzmann code, and scale it back to the starting redshift using the Newtonian, scale independent {radiationless} growth factor {$D_{\rm N}(k,z)=D_{\rm N}(z)$:
\begin{equation}
T_i(\tau(z_i), \vec k) =  \left(\frac{D_{\rm N}(z_i)}{D_{\rm N}(z=0)}\right) T_i(\tau(z=0), \vec k)~.
\end{equation}
Since at redshift zero, the baryons have fallen into the potential of dark-matter, the rescaling approach provides baryons with practically the same spectrum as dark matter even at the initial redshift, making this approach very similar to the total-matter approximation above.

\subsection{Longitudinal gauge velocities}\label{subsec:gauges}
This approximation consists in computing displacements and peculiar velocities through the power spectrum in the synchronous and longitudinal gauge, respectively. The longitudinal gauge is defined by choosing $B_i = h^T_{ij} = 0$ {in Eq.~\eqref{eq:anyGauge}}. {When in addition to this gauge choice, the solution for a certain matter content dictates $\dot H_L = 0$, such as in a pure dust Universe}, the velocities on sub-Hubble scales in this gauge correspond to those of the N-body gauge. However, at the linear level, at early times the residual radiation perturbations cause a small time dependence in $H_L$. At late times a cosmological constant cause an additional time dependence in $H_L$. These effects are small, as is evident from the solid line for ``Longitudinal-gauge velocities'' in Fig~\ref{fig:veloResults}: linear theory predicts only a minimal error at intermediate to large scales. {In summary, this approximation assigns the wrong velocities on scales at and beyond the Hubble radius, which is inside the simulation volume at initial redshifts. Note that this approach differs from the fiducial model only at the level of velocities. The displacements are identical to the fiducial approach.

\subsection{Offset grids}\label{subsec:offsets}
In order to avoid spurious two-particles interactions, caused by placing baryons and dark matter particles at identical positions before convolving with their respective transfer functions, one needs to offset the baryons by half the inter-particle distance, on a staggered grid \cite[see][for the impact of setting the ICs using grid or glass files]{Yoshida:2003sy}. Using a Fourier transform, this can be done exactly by, \citep{Angulo:2013qp}, 
\begin{align}
f(\vec x + \vec y) = \int \frac{d^3k}{(2\pi)^3} e^{-i\vec k \cdot \vec x} e^{-i\vec k \cdot \vec y} f_{\vec k},
\end{align}
such that the values of the field in a grid offset by a constant delta in each dimension, is obtained by adding a constant phase to the field in fourier space:
\begin{align}
f_{\vec k} \rightarrow f_{\vec k'} = e^{-i\vec k \cdot \vec y} f_{\vec k}. 
\end{align}

Many existing simulations do not take this phase shift into account, and provide baryons and dark matter with identical displacements and velocities while placing them on offset positions. {Note that the error associated to the way mode amplitudes are set on the regular grid in Fourier space has been recently studied in \cite{2016arXiv161004862F}.} 

When one uses the phase shift as described here, in practice one interpolates a grid of $N^3$ points on an offset grid of $N^3$ intermediate points. This means that the discrete set of points, actually is capable of representing power at higher frequencies than a single grid can. The interpolation implies that this higher-frequency power is artificially cutoff by a tophat filter. One may argue that is would be more realistic to actually generate the initial power on a $(2N)^3$ grid, and sample the two staggered $N^3$ grids on that higher resolution grid, such that no power goes missing. One downside is that one needs to be able to fit the $(2N)^3$ grid in memory. The actual effects of this interpolation and power cutoff are left open for further research.

\section{Errors at the initial time}
\label{sec:errors_ICs}
At the time of setting initial conditions, one disposes of the practically exact Boltzmann solutions for the linear power spectra. One can hence straightforwardly quantify the errors made in the spectra of the species by comparing the effective transfer functions used for each species to the reference transfer function, as show in Fig.~\ref{fig:compLinear}. Clearly, the various approaches introduce an error of several percent in both baryonic and dark-matter power spectra. It is important to point out that this applies to the spectra for both densities (which give displacements) {\em and} peculiar velocities. Notably, baryons have a different scale dependent growth than dark matter, and hence will have varying directions~\citep{Valkenburg:2015dsa}.

The approach of `wrong offsets'~(\ref{subsec:offsets}) formally gives both baryons and dark matter the correct power spectrum. Yet, in this setup the total matter power spectrum will be wrong, as the addition of waves with the wrong phase leads to cancellations in the total density distribution. However, in the limit of infinite resolution, the error in this approach tends to zero, as the relative offset between the dark-matter and baryon grids tends to zero.

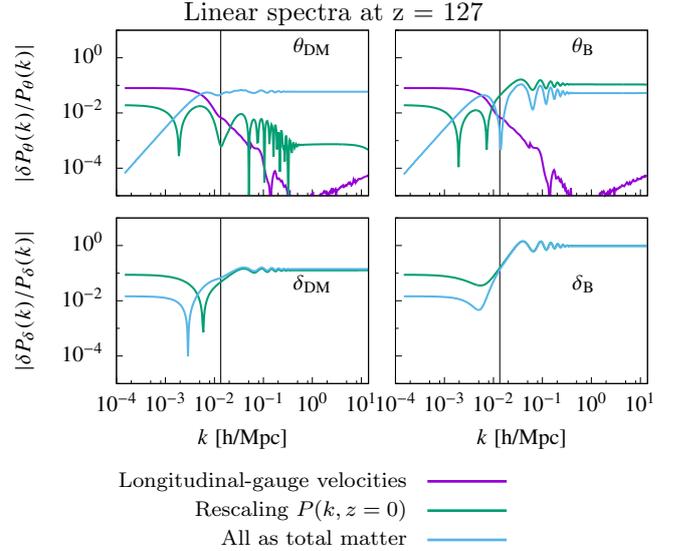
\begin{figure}
\begin{center}
{\large Linear spectra at z = 127}
\includegraphics[width=\columnwidth]{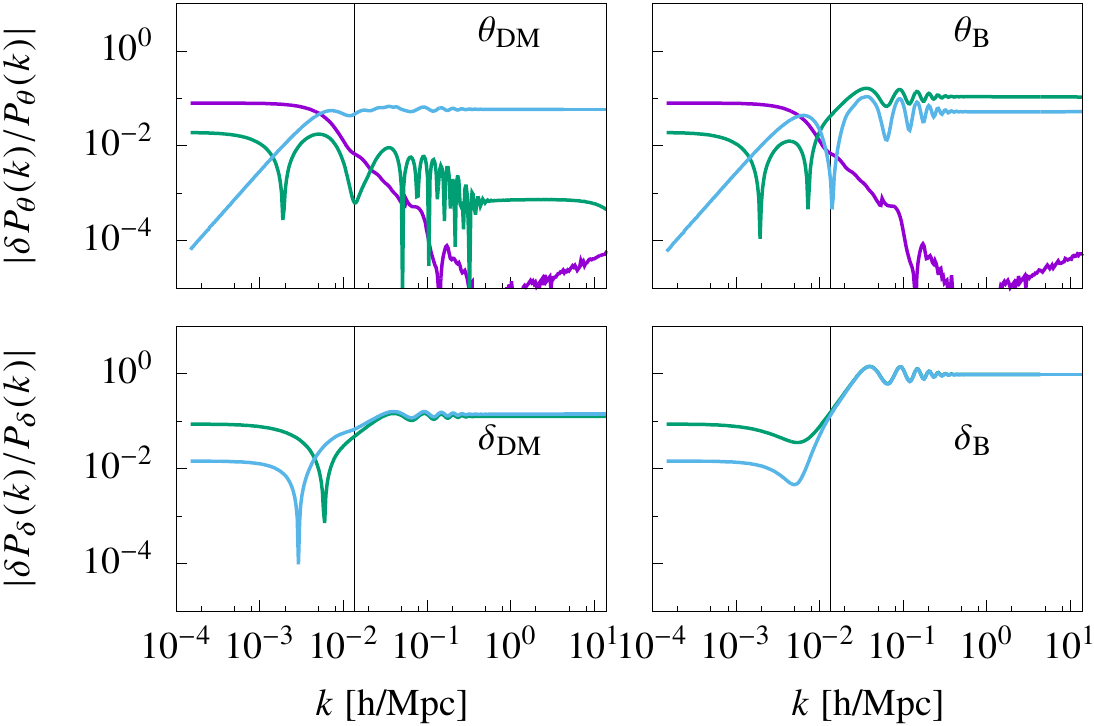}
\setlength{\unitlength}{0.0500bp}%
\begin{picture}(0.00,1100.00)%
  \csname LTb\endcsname%
  \put(548,886){\makebox(0,0)[r]{\strut{}Longitudinal-gauge velocities}}%
  \csname LTb\endcsname%
  \put(548,666){\makebox(0,0)[r]{\strut{}Rescaling $P(k,z=0)$}}%
  \csname LTb\endcsname%
  \put(548,446){\makebox(0,0)[r]{\strut{}All as total matter}}%
\put(700,420){\includegraphics{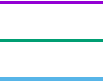}}%
\end{picture}%
\end{center}
\vspace*{-10mm}
\caption{{Comparison between the velocity (upper row) and density (bottom row) power spectra of dark matter (left column) and baryons (right column) between the initial conditions generated according to each approximation and the prediction of {\sc CLASS} at $z=127$.}
Absence of a line means that the error is zero. The vertical black line indicates the wavenumber that corresponds to the Hubble parameter  in comoving coordinates, $k_{\rm Hubble} = 2\pi/aH$. Note that ``Rescaling $P(k,z=0)$'' implies ignoring radiation at the background level, such that redshift and growth factor do not map properly to our reference model.}\label{fig:compLinear}
\end{figure}

\section{Resulting errors at end of simulation}
\label{sec:errors}

{In this section we compare the result of a hydrodynamic simulation run with the ICs generated without assumptions (our fiducial model) versus the results of different hydrodynamic simulations whose ICs are generated employing the above approximations.}

\subsection{Setup}
\begin{figure}
\begin{center}
{\large Reference simulation}
\includegraphics[width=\columnwidth]{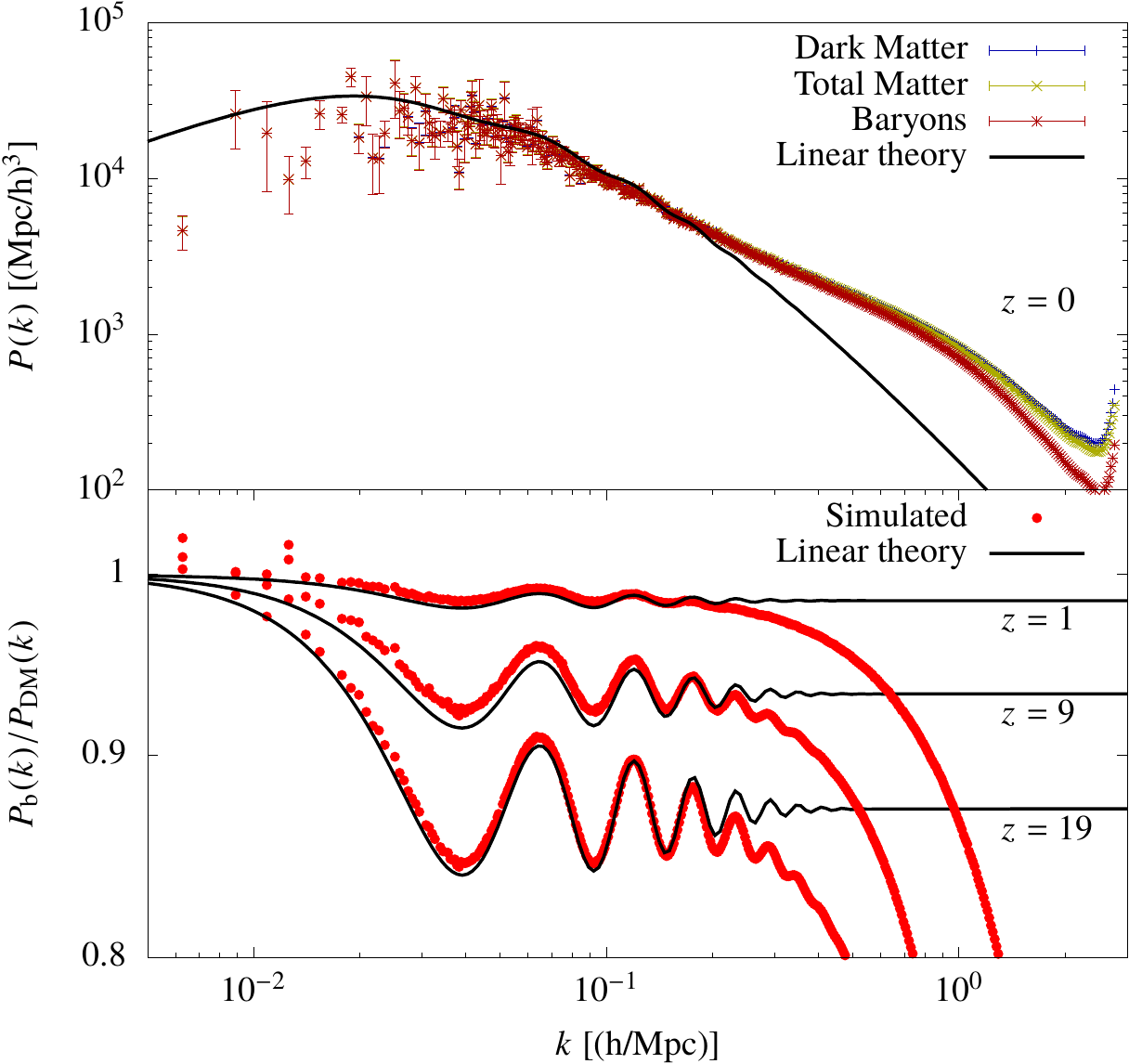}
\end{center}
\caption{{\em Top:} The power spectra of baryons, dark matter and their weighted sum at redshift zero. The dark-matter spectrum is hidden by the total matter spectrum. The baryon, dark-matter and total-matter spectra from linear theory are indistinguishable in this logarithmic plot. {The error bars indicate the variance in a bin, stemming from the finite boxsize. Note that the variance itself is subject to the generated random numbers, such that the small variance in the bin for smallest k is a random fluke in a bin with very few modes.} {\em Bottom:} The ratio of the baryon spectrum to the dark-matter spectrum at various redshifts (compare to Fig. 1 in~\citep{Angulo:2013qp}), demonstrating that our choice of TreePM parameters reproduces linear theory reasonably well.  }
\label{fig:refSimDemo}
\end{figure}

{The initial conditions are generated by placing dark matter and gas particles into two different regular grids offset in each cartesian coordinate by half the grid size, and then displacing the particles and assigning those peculiar velocitties according to the Zel'dovich approximation with spectra from the N-body gauge. {Note that in all the runs} we have taken into account the scale-dependent growth rate present in the simulations\footnote{{We notice that in contrast with N-body simulations of a single matter fluid, in simulations involving both dark matter and gas the growth factors/rates and scale-dependent as in the case of massive neutrino cosmologies \citep{Zennaro:2016nqo}.}} (see Fig. \ref{fig:growth_rate}). We obtained the linear matter power spectrum and the transfer functions by using {\sc CLASS}. We have used the code FalconIC~\citep{Valkenburg:2015dsa}, publicly available at \url{http://falconb.org}, to generate the simulation ICs.}

In our simulations we followed the evolution of $512^3$ dark matter plus $512^3$ gas particles in a periodic box of $1~h^{-1}$Gpc using {\sc Gadget-III} \citep{Springel:2005mi}. In our fiducial run we also account for the contribution of radiation to the Hubble function. The softening length of each particle type is set to $1/40$ of the mean inter-particle distance, on a gravity mesh of $512^3$ cells, with the long-range/short-range force-split (``ASMTH'') set to 0.125 times a cell width. The values of the cosmological parameters are: $\Omega_{\rm m}=0.3175$, $\Omega_\Lambda=0.6825$, $\Omega_{\rm b}=0.049$, $n_s=0.96$, $h=0.67$ and $\sigma_8=0.83$, in excellent agreement with the latest Planck results \citep{Planck_15}.

In the upper panel of Fig. \ref{fig:refSimDemo} we show the dark matter, baryon and total matter power spectra, together with linear theory prediction by {\sc CLASS}. As expected, we find that all species have basically the same clustering on large-scales, in agreement with linear theory. In the bottom panel of that figure we display the ratio between the power spectra of baryons and dark matter at different redshifts, together with the prediction by linear theory. We find an excellent agreement between the results of our simulations and linear theory, with subpercent offsets likely due to the resolution of our simulations. 

Since in this paper we are just interested on the impact of the different approximations used when generating initial conditions we have employed a simplified hydrodynamic scheme were the gas only is subject to adiabatic cooling, i.e. we do not consider radiative cooling, star formation and supernova/AGN feedback.

\subsection{Test cases}
{Next to our {fiducial} simulation, we set up four different initial conditions, identical to the reference model (including the random seed), apart from the initial power spectra of the species. That is, we generated initial conditions with each of the four approximations listed in sections~\ref{subsec:multispecies},~\ref{subsec:veryoldway},~\ref{subsec:gauges}~and~\ref{subsec:offsets}. We ran {hydrodynamic} simulations off of each of these four initial conditions, in order to compare the final result to the reference model.}

\subsection{Results}

{For each simulation we have computed the gas, dark matter and total matter density and velocity power spectrum at $z=0$. In Figs. \ref{fig:veloResults}~and~\ref{fig:densResults} we display the relative difference between the power spectra from the simulations whose initial conditions have been generated using a different approximation to the one where the ICs were created without approximations. The bottom panels of those figures show a zoom into the different curves together with the prediction from linear theory at $z=0$. We find that linear theory is no good guidance for the final error in the simulation caused by the initial conditions.}

As expected, we find that different approximations lead to different biases. By setting velocity amplitudes using the longitudinal-gauge we find that the error on most of the scales proved by our simulations to be negligible, although differences reach the percent level on large-scales and increase very rapidly, suggesting that this approximation can lead to significant effect on very large box size simulations. The largest error we encounter arises by using the same, total matter, transfer functions when setting up the initial conditions of both dark matter and gas. We find that the magnitude of the effect can be as large at $\sim10\%$ in the velocity power spectrum. By employing the rescaling procedure we find that the error locates between the previous two ones, with a sub-percent magnitude in most of the cases. As expected, the errors associated to using the wrong offset when generating the ICs it is only important on small scales.

Even though the rescaling procedure leads to an internally self-consistent setup, whose linear redshift-0 spectrum is identical to the linear fiducial spectrum, the nonlinear output of the simulation is different. How the difference exactly arises is a matter for further research, but we can point out that both setups model slightly different cosmologies, with different expansions rates. Moreover, the power spectra of baryons and dark matter employed to generate the ICs in both situations are different: at $z=0$ the linear power spectra of both components are very similar, and rescaling those back will not make them different, while the power spectra of each species from {\sc CLASS} at $z=127$ is significantly different to each other (the fact that power spectra are different at at $z=127$ can be seen in Fig. \ref{fig:compLinear}).

It is interesting to point out that for approximations ~\ref{subsec:gauges}~and~\ref{subsec:offsets}, there is initially no error in the density spectra of the individual species, where for the latter approximation also no error in the velocity spectra exists. However, the final spectra do contain an error in all quantities for all species. 

\begin{figure*}
\includegraphics[width=2\columnwidth]{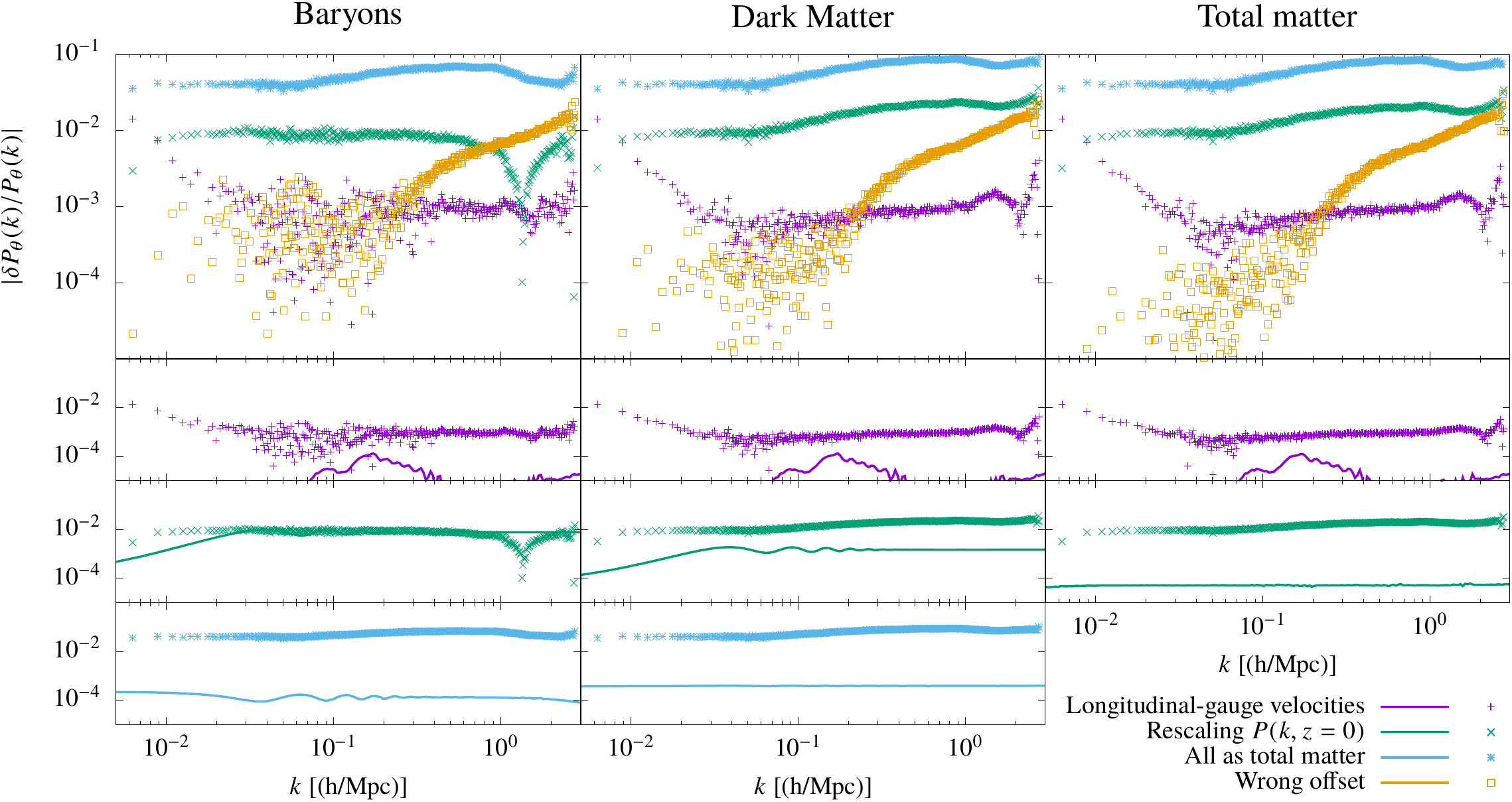}
\caption{The deviation of the velocity-divergence power spectra of simulations with flawed initial conditions as described in~\ref{subsec:multispecies},~\ref{subsec:veryoldway},~\ref{subsec:gauges},~\ref{subsec:offsets} relative to the reference simulation. {\em Top row:} A comparison of the various simulations. {\em Other rows:} Each of the simulations deviation {(data points)} compared linear perturbation theory {(solid lines)} at redshift $z=0$, where applicable: approximations whose linear $z=0$-spectrum predicts zero error are left out. The error at the initial redshift $z=127$ survives in the simulation down to $z=0$, and does not decay as opposed to the linear perturbation theory prediction. }
\label{fig:veloResults}
\end{figure*}

\begin{figure*}
\includegraphics[width=2\columnwidth]{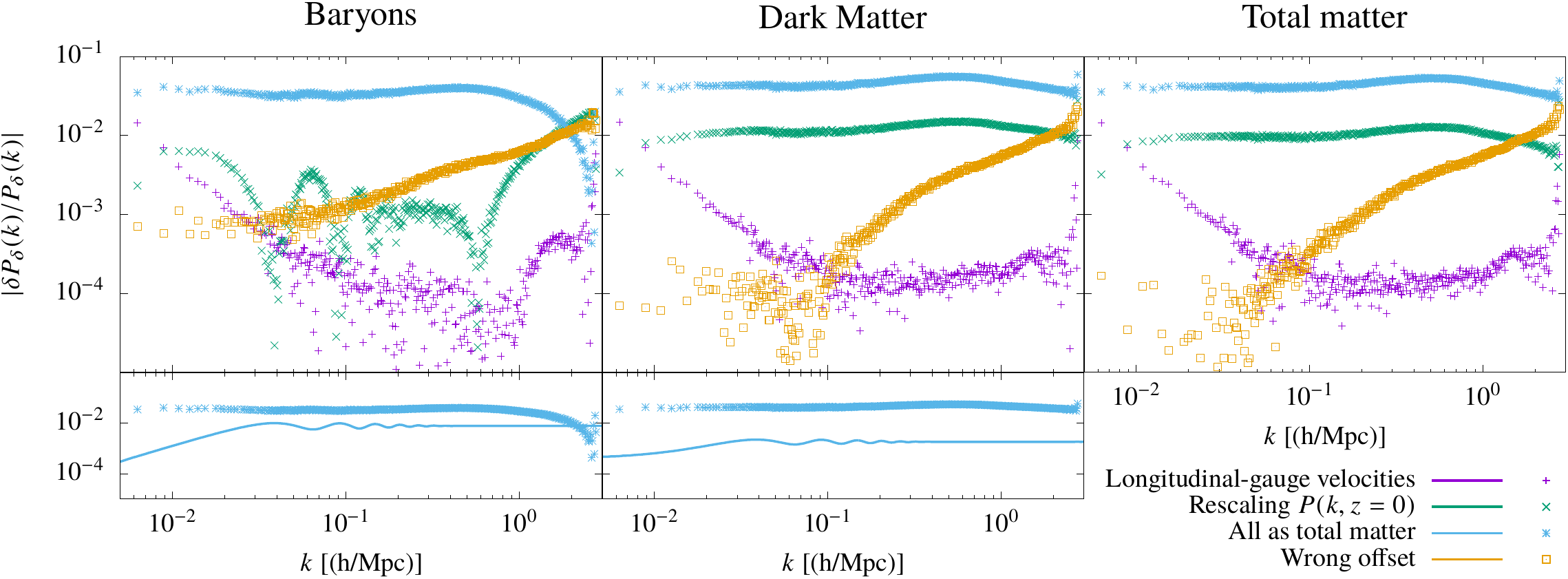}
\caption{The deviation of the {density} power spectra of simulations with flawed initial conditions as described in~\ref{subsec:multispecies},~\ref{subsec:veryoldway},~\ref{subsec:gauges},~\ref{subsec:offsets} relative to the reference simulation. {\em Top row:} A comparison of the various simulations. {\em Other rows:} Each of the simulations deviation compared linear perturbation theory at redshift $z=0$, where applicable: approximations whose linear $z=0$-spectrum predicts zero error are left out.  }
\label{fig:densResults}
\end{figure*}

\section{Conclusions and discussions}
\label{sec:conclusions}

{Upcoming cosmological surveys will collect enough data to allow subpercent constraints on the value of the cosmological parameters, that will shed light on the several aspects such as the nature of dark energy and the neutrino masses. The constraints on the value of the cosmological parameters are derived by comparing observational data against theoretical predictions. For the latter, numerical simulations play an increasingly important role, since they provide results into the fully non-linear regime and as their own nature accounts for all complications inherent to cosmology such as redshift-space distortions, halo/galaxy bias and so on. }

{A very import step involved in the process of running a numerical simulation is to generate its initial conditions. It is a standard practice to make use of different approximations when setting up the ICs of a numerical simulations. The purpose of this paper is to quantify the error induced by those approximations when running hydrodynamic simulations with both dark matter and gas particles.}

{We find that for simulations with boxes not large enough, where radiation perturbations become important, the correct way to generate ICs is to use the transfer functions from Boltzmann solvers such as {\sc CAMB} or {\sc CLASS} and account for both the scale-dependence growth factors/rates present in the case of several fluids. We also find that accounting for the contribution of radiation to the Hubble function becomes important at high-redshift and can significantly change the results on those epochs.}

Using any of the wrong approximations leads to an inconsistent setup, with the wrong boundary conditions for the correct set of equations to be solved. Not surprisingly, as nonlinearities grow, the initial error does {\em not} decay, rather it survives until redshift zero. The error in large baryon--dark-matter simulations, such as the Illustris simulation~\citep{Vogelsberger:2014dza} and the Eagle suite~\citep{Schaye:2014tpa}, is {\em at least} several percent in the spectra of the total matter, dark matter and baryons, as these simulations make three of the unnecessary approximations at once (rescaling the total matter power spectrum from $z=0$ to the initial redshift, assigning that total matter power to both species, without correcting for the offset grids).

We however emphasize that even if the initial conditions have been set up properly, numerical artifacts can couple small scale power from two different fluids \citep{Angulo:2013qp} erasing the memory from the initial conditions and yielding results in disagreement with the expectation of linear theory on large scales. {Secondly, }unless the simulation code is able to account for radiation perturbations, following the Newtonian dynamics of a set of particles whose ICs have been set up using the output of a Boltzmann code at high-redshift, will induce clustering differences on large-scales with respect to the Boltzmann code. A way to resolve this issue would be, following \cite{Zennaro:2016nqo}, to write down the equations governing the Newtonian evolution of a system comprised by two fluids and extract from that system the growth factors/rates and use those to "rescale" the $z=0$ power spectrum from the Boltzmann code, or to add radiation perturbations to the construction of the gravitational potential at each time step~\citep{Brandbyge:2016raj}. This will guarantee the correct clustering amplitude on linear scales and will reproduce the correct clustering of the Boltzmann code at low redshift. We plan to investigate this in a separate paper.

Whether one should set the initial conditions using the method described in this paper or using rescaling techniques, depends on the type of output desired. While the former will produce outputs at high-redshift (low-redshift) with clustering properties in agreement (disagreement) with Boltzmann solvers, the situation is the opposite when using rescaling techniques. We believe the method presented here is well suited for small box size simulations down to $z=0$ (such as Illustris of Eagle) of for large box size simulations at high-redshift such as those to study the Epoch or Reionization.

\section*{Acknowledgements}
{The work of FVN has been supported by the Simons Foundation and by the by the ERC Starting Grant ``cosmoIGM''.} The authors thank Matteo Zennaro, Guido d'Amico, Marco Baldi, Oliver Hahn and Raul Angulo for useful discussions. The simulations were performed on the Novamaris Cluster at the Instituut-Lorentz, University of Leiden and the thqcd2 and hpc-qcd PC clusters at CERN.

\bibliographystyle{mnras}
\bibliography{refs}

\end{document}